# On Limitations of the Witness Configuration Method for Geometric Constraint Solving in CAD Modeling

Qiang Zou*, Hsi-Yung Feng

The University of British Columbia
Vancouver, BC
Canada V6T 1Z4

**Abstract**

This paper presents discussions on the limitations of the witness configuration method. These limitations have rarely been reported in previous studies. The witness configuration method is a very recent approach for geometric constraint solving that is of critical importance for modern computer-aided design systems. The witness configuration method is likely the most promising method to solve satisfactorily the challenges of geometric constraint solving. This method, in the current form, is however found to be limited for the three essential tasks in geometric constraint solving. Examples are given to validate this work's statements on these limitations.

**1. Introduction**

The notion of computer-aided design (CAD) may date back to the 1960s [1,2]. Nevertheless, it was not until the introduction of parametric CAD modelers in the late 1980s [3] that CAD was recognized by the user community. The main practical benefit of parametric CAD lies in geometry reuse, which allows users to attain geometry variants through tuning parameters associated with the geometry. While parametric CAD is not a single technique but a set of techniques, common to all parametric techniques is associativity that allows local parametric changes to propagate automatically. The associativity is often implemented as a system of geometric constraints relating geometric entities in the model; the automatic change propagation is done by re-evaluating the geometric constraint system, which is referred to as geometric constraint solving.

A geometric constraint system (GCS) involves a set of geometric entities like point, line, plane, and cylinder, as well as a set of geometric constraints between the entities such as distance and angle. These constraints can usually be translated to algebraic equations whose variables are the parameters of the participating geometric entities. Solving a GCS refers to the process of attaining a solution (or solutions) to the corresponding algebraic system. Since the solving is understood in a bigger picture of geometry, a solution is a valid instantiation of the geometric entities such that all the constraints are satisfied.

Geometric constraint solving may roughly consist of two phases: non-well-constraint resolution and real solving. The GCS given through the user's specification could contain under- and over-constrained parts. Such parts should be correctly detected and then resolved before proceeding to the solving phase. *One governing issue is to have generic criteria to characterize constraint states (under-, well-, and over-constrained).* It is not until we have these criteria that detection of under-constrained and over-constrained parts can be made possible. *Another issue is to optimally detect these parts for better resolution guidance provided to the user, which involves attaining minimal over-constrained parts and maximal well-constrained parts.* The major issue in the solving phase is efficiency. GCS's in CAD could be very large, and solving such systems is computationally challenging. Currently, researchers address this issue through a pre-processing step of decomposing a given (well-constrained) GCS into small well-constrained subsystems, whose solutions can be assembled to attain the overall solution [4]. *Again the criteria of characterizing constraint states are building blocks for the decomposition.* In addition, the overall solving efficiency is governed by the efficiency of solving the largest subsystem [4]. *Hence, the optimal decomposition plan would minimize the size of the largest subsystem*. In summary, there are three basic issues in geometric constraint solving, as listed below. In the authors' opinion, the first two issues are primary and the last one is secondary since the efficiency issue, if we do not solve it, will be solved by time, a result of the exponential growth in computing power (Moore's law) or advances of quantum computing techniques.

---

* Corresponding author. Tel.: +1-778-251-0741. E-mail: john.qiangzou@gmail.com

I. Effective criteria of characterizing under-, well-, and over-constraint states;
II. Optimal decomposition of a non-well-constrained constraint system into minimal over-constrained parts and maximal well-constrained parts for efficient GCS resolution; and
III. Optimal decomposition of a well-constrained constraint system into minimal well-constrained parts for efficient GCS solving.

Several research directions have been presented in the literature to address these issues. The very recent witness configuration method (VCM) is likely the most promising method, especially in the context of the very recent direct modeling CAD paradigm [5,6]; other methods have proven to have inherent limitations and are often used beyond their validity. Nevertheless, VCM's current development is found to be insufficient. This work discusses what has already been made possible for VCM, what developments can be expected in the near future, and which areas remain problematic, in terms of the three issues.

The remainder of this paper is organized as follows. Section 2 begins with a brief, non-exhaustive summary of the historical development of geometric constraint solving, showing where VCM starts from; then a detailed review of VCM is to be given to note VCM's state of the art. The following Limitations section consists of two parts: the first part discusses VCM's basic idea; and the second part presents how the existing work deals with the three issues, as well as their limitations. The last section concludes this paper.

## 2. Related work

Use of constraints in CAD can date back to the very first CAD prototype SketchPad [7], where constraints were used to precisely define 2D shapes. At that time, no particular attention was paid to the problem of solving constraint systems, and the solving methods used were a direct application of general-purpose numeric methods like the Newton–Raphson method. With the development of CAD techniques, especially to the time of parametric CAD modeling, constraint systems became large and complex, introducing the issues of how to characterize the constraint state of the GCS specified by the user and how to solve them efficiently. In the last three decades, a good number of publications related to these issues have been presented, refer to [8] for a thorough review of them. The ideas presented may be classified into four categories: solving-based, logic-based, graph-based, and perturbation-based.

The solving-based methods are the (conceptually) simplest, which analyze the constrained state of a GCS by directly solving the GCS through symbolic methods (e.g., Grobner bases, Wu-Ritt triangulation methods) or numerical methods (e.g., Newton-Raphson, homotopy continuation methods) [8]. These methods do not consider the other issue of GCS solving efficiency. In fact, they are rarely used in real practice due to the high computational load.

The logic-based methods like [9,10] develop a set of geometric theorems and derivation rules and then, based on them, check if the GCS can be logically derived. If so, the GCS is well-constrained; if there are extra constraints, it is over-constrained; otherwise, it is under-constrained. In addition, this derivation process can find well-constrained subsystems in a given well-constrained GCS, and thus can be used to attain a decomposition plan for the GCS to attain efficient solving. The logic-based method is essentially an axiomatization of all possible constraint systems, which is of great mathematical elegance. However, the library of the geometric theorems and derivation rules developed is far from being sufficient for practical usages.

The graph-based methods handle the GCS in an indirect way by converting a GCS to a graph structure and then studying this graph instead of the original GCS. There are two lines of development. The first one tries to recognize pre-crafted graph patterns (corresponding to known shapes) in a GCS. This line was pioneered by Owen [11] and much improved by [12–14] in terms of the size of the pattern library. The second line compares degrees of freedom (DOFs) of the geometry with degrees of restriction of the GCS. This line was initialized by Bardord [15] and Serrano [16], and detailed in [17–19]. In 2001, the two lines were unified under a framework proposed by Hoffmann et al. [4]. Ever since, there have still been some good progress, yet the foundations remain unchanged. Although graph-based methods could be effective in many scenarios, a known limitation is the incapability to handle constraint dependencies (except for the simplest structural dependencies) [20]. The reason is that the graph representation only captures the combinatorial information of the GCS and discards the geometric information.

To overcome the limitations of graph-based methods, VCM was proposed [20], which basically revisited and improved the previous infinitesimal rigid theory [21]. This method examines how the constraint equations behave under the infinitesimal perturbations made to the constraint equation variables, whose relationship is described by the associated Jacobian matrix of the constraint equations. That is, the dependent rows of this matrix characterize the (structural and non-structural) constraint dependencies and the kernel of this matrix gives the DOFs of the GCS [22,23]. A successful application of this method requires the Jacobian to be evaluated at a carefully selected point called witness configuration, which has satisfied all the constraints with degenerate in the GCS, for example, parallel and incident constraints [24]. It is for this reason that this method is named VCM. This method has been successfully applied to some real problems such as [25,26]. At present, it has been proven that all constraint dependencies (i.e., over-constraint) can be detected using this method [23]. Nevertheless, according to the authors' investigation, the developed criteria to characterize well-constraint or under-constraint are not that effective and may give wrong results in some situations (as will be presented shortly). As for Issue II, the developed methods [22,23,25,26] to detect the minimal over-constrained parts and maximal well-constrained parts in a GCS are also insufficient: greedy algorithms are used to find such parts, but it is found that these algorithms cannot always do so correctly. With regard to Issue III, there seems to be a lack of published research work on it, to our knowledge. There may be unpublished work undergoing development.

### 3. Limitations of VCM

As already noted, a given GCS can be expressed as a system of algebraic equations, denoted as $F(X) = 0$ where $X$ represents the parameters of the participating geometric entities. Through examining the solution(s) of this system, notions of under-, well-, and over-constraint can be formally defined as follows, which is the base of the discussion in this work.

**Definition 1.** Let $F(X) = 0$ be the constraint equations of a GCS, and $S$ the solution space. The GCS is consistently constrained if $S \neq \emptyset$, and inconsistently constrained otherwise.

**Definition 2.** Let $F(X) = 0$ represent a consistently constrained GCS, and $S$ the solution space. The GCS is under-constrained if the cardinality $|S|$ is infinite.

**Definition 3.** Let $F(X) = 0$ represent a consistently constrained GCS, and $S$ the solution space. The GCS is consistently over-constrained if there exists a subset of the GCS such that the corresponding reduced equations have the same solution space as $S$.

**Definition 4.** A GCS is over-constrained if it is inconsistently constrained or consistently over-constrained.

**Definition 5.** A GCS is well-constrained if it is neither under-constrained nor over-constrained.

*3.1. A brief introduction*

VCM is essentially a bunch of criteria to characterize constraint states. It attains these criteria by examining how the constraint equations behave under infinitesimal perturbations made to the equation variables. That is, VCM applies a small perturbation $\Delta X$ to $X$ and examines the response $\Delta F = F(X + \Delta X) - F(X)$ of the constraint equations, which are related by (the first order approximation):

$$\Delta F = J(X) \cdot \Delta X + O(\|\Delta X\|^2) \tag{1}$$

where $J(X)$ is the Jacobian matrix evaluated at point $X$. The main result of VCM is that if $J$ is evaluated at a carefully selected point called witness configuration, the constraint dependencies (i.e., over-constraint) between the equations are described by the linearly dependent rows of the Jacobian matrix, i.e., the vectors in the kernel (or null space) of $J^T$. Under-constraint is closely related to the free perturbations that do not violate existing constraints, i.e., $\Delta F = J \cdot \Delta X = 0$; that is, under-constraint is described by the vectors in the kernel of $J$. A configuration is deemed a witness configuration if certain types of constraints in the given model has already been satisfied by the configuration under perturbation, i.e., $X$, see [24] for generation of such configurations. A detailed discussion of witness configurations and the necessity of them for VCM to work well can be found in [22,23].

The formal criteria of under-, well- and over-constraint given by VCM are as follows [22]. A model is over-constrained if $Kernel(J^T) \neq \emptyset$, where $Kernel(\cdot)$ denotes the kernel of a matrix: $Kernel(A) = \{x \mid Ax = 0\}$. This criterion can be rewritten as:

$$Rank(J) - RowSize(J) < 0 \qquad (2)$$

A model is under-constrained if $Dim(Kernel(J)) > 6$, where $Dim(\cdot)$ is to get the dimension of a linear space and the number 6 represents the 6 rigid-body transformations. This criterion can be rewritten as:

$$Dim(Kernel(J)) = ColumnSize(J) - Rank(J) > 6 \qquad (3)$$

using the rank-nullity theorem [27]. Following immediately, the criterion for a well-constrained GCS is given as:

$$\begin{aligned} Rank(J) - RowSize(J) &= 0 \\ ColumnSize(J) - Rank(J) &= 6 \end{aligned} \qquad (4)$$

These criteria were accepted once, but were later proven wrong. The failure is due to the assumption that a well-constrained GCS admits exactly 6 DOFs, which is seen to be not true in many real cases [28]. Fig. 1 shows such a failure example[1]. To address this issue, the recent VCM replace the fixed number 6 with a variable called degree of rigidity (DOR) [23]. The DOR notion may be first presented in [28], and then was directly applied to VCM in [23]. This notion essentially translates the 6 rigid-body transformations (3 axial translations and 3 axial rotations) to changes made to the parameters of the participating geometric entities, where these changes are stored as 6 vectors; then DOR of a GCS is given by the number of independent vectors among the 6 vectors, see Section 3 of [23] for details. Consider, again, the example in Fig. 1. Its DOR is 5; then the numbers $ColumnSize(J) - Rank(J)$ and DOR are matched. This means that the constraint state of this example is characterized correctly with the help of the DOR notion.

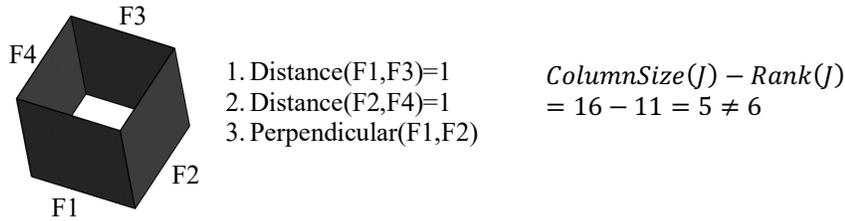

Figure 1  A failure example: a well-constrained model but incorrectly characterized by VCM.

*3.2. Limitation I*

The first limitation to be discussed in this subsection concerns with the ineffectiveness of VCM's criteria for characterizing constraint states. Although enhanced with the DOR notion, there is still no rigorous proof for VCM. The introduction of the DOR notion is more like a patch for fixing the bugs encountered in the application of VCM than a systematic treatment of VCM's limitations. In fact, none of the key publications on VCM, say [20,22,23], come with guarantees or with clearly stated limitations. Effectiveness in terms of constraint state characterization was often over-claimed in previous publications but was negated in later publications, which is the case for the introduction of the DOR notion to VCM.

One limitation encountered by the authors in applying VCM is that the representation scheme used to describe the participating geometric entities has a significant effect on VCM's effectiveness. Consider, for example, the four-plane configuration in Fig. 1, and change the original representation scheme to the following one: a plane is described with a point $p \in R^3$ on the plane and the normal $n \in S^2$ of the plane. The expression $ColumnSize(J) -$

---

[1] In this example, a plane is described with the tuple $(a, b, c, d)$ and the additonal constraint $a^2 + b^2 + c^2 = 1$ (the plane equation is $ax + by + cz + d = 0$). The equations of the involved constraints are as follows: (1) collinearity of two normal or direction vectors $v_1, v_2$ is represented with $v_1 + tv_2 = 0$ where $t$ is a scalar unknown; and (2) point-to-point distance and vector angle are represented with dot products.

$Rank(J)$ becomes $24 - 13 = 11$, and DOR of this example becomes 6, which leads to a mismatch and a failed constraint state characterization. From this example, we can conclude that different representation schemes could result in different numbers of the terms $ColumnSize(J)$, $Rank(J)$, and DOR, and consequently could lead to opposite characterization results for a same GCS. Fig. 2 gives one more example about such a situation, and Table 1 summaries the characterization results for this example and the one in Fig. 1. Representation #1 refers to the original representation scheme; Representation #2 refers to the scheme we just mentioned above; and the representation scheme used to describe the lines of the example in Fig. 2 is: a line is represented by a point $p \in R^3$ on the line and the direction $d \in S^2$ of the line. Based on the results in Table 1, it is safe to say that the current VCM (with the DOR notion) is not as effective as we thought.

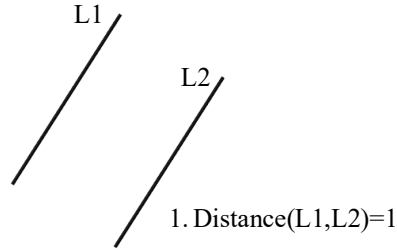

Figure 2  Another failure example: a well-constrained model but incorrectly characterized by VCM.

Table 1  Characterization results for the plane and line examples.

|  | VCM Without DOR Notion | | | | VCM With DOR Notion | | | |
| --- | --- | --- | --- | --- | --- | --- | --- | --- |
|  | $ColumnSize(J)$ | $Rank(J)$ | DOR | Matched? | $ColumnSize(J)$ | $Rank(J)$ | DOR | Matched? |
| The plane example (Representation #1) | 16 | 11 | 6 | ✗ | 16 | 11 | 5 | ✓ |
| The plane example (Representation #2) | 24 | 13 | 6 | ✗ | 24 | 13 | 6 | ✗ |
| The line example | 12 | 5 | 6 | ✗ | 12 | 5 | 6 | ✗ |

The above limitation only represents one limitation found during the authors' application of VCM. Are there other hidden limitations? At present, the authors do not have a good answer to this question. There are good reasons that it may be better for us to take the negative side: existing work has over-claimed the effectiveness of VCM twice, of which one is before the introduction of DOR, and the other before this work. Despite of these limitations (found ones and future ones), VCM is still a promising (maybe the most promising) method for handling the task of constraint state characterization (other known methods are less effective than VCM and have inherent limitations). Further development of VCM is necessary to prefect VCM (for constraint state characterization). Two improvement directions maybe taken to achieve the goal: (1) have a comprehensive analysis of VCM's capabilities and limitations, and then device new (add-on) methods to specifically handle the limitations; and (2) propose a variant of VCM for which a rigorous proof is achievable. The authors have presented a systematic method in [29] to address Limitation I, following the latter direction.

*3.3. Limitation II*

This subsection and the next discuss the limitations of VCM for dealing with the second issue in geometric constraint solving. To repeat, the issue concerns with the optimal decomposition of a non-well-constrained constraint system into the minimal over-constrained parts and the maximal well-constrained parts. The limitation this subsection focuses on is the ineffectiveness of the current methods on minimal over-constrained part detection.

Over-constrained parts in a non-well-constrained GCS could take the form of a group of dependent constraints, or the form of a group of geometric entities whose associated constraints are more than needed. These two forms are equivalent; and this work defines an over-constrained part as a group of dependent constraints. The task of minimal over-constrained part detection is to minimize the size of each over-constrained part in the GCS.

Existing methods to accomplish the task share the same idea of using greedy methods [22,23,25,26]. Conceptually, they first find a maximal subset of independent constraints in the given GCS, then decide the dependencies of the rest constraints on these independent constraints in a one-by-one manner, which can be understood using the simple example in Eq. 5. To simplify the indexing of these equations, they are numbered E1-E5 sequentially. The existing methods start from a seed equation, e.g., E1, then iterate through all equations to greedily find a maximal subset of independent equations; for this case, it will be {E1, E2, E3}. Subsequently, the dependencies of the rest constraints with the constraints already presented in the subset will be checked. For example, equation E4 is dependent with all the three constraints in the subset, and the same for equation E5. According to these dependencies, the methods will output two minimal dependency groups: {E1, E2, E3, E4} and {E1, E2, E3, E5}.

$$\begin{aligned} x + y + z &= 0 \\ 2x + y + z &= 1 \\ 3x + 2y + z &= 1 \\ x + 2y + 3z &= 1 \\ 2x + 4y + 3z &= 2 \end{aligned} \quad (5)$$

The above strategy may fail to detect the minimal over-constrained parts in a GCS because the maximal independent subset of a constraint system is not unique; this non-uniqueness could lead to wrong detection results. Consider the example in Eq. 6 (a slight modification of Eq. 5). The existing methods will give two "minimal" dependency groups: {E1, E2, E3, E4} and {E1, E2, E3, E5}. However, it is easy to see that there is another dependency group with smaller size: {E4, E5}.

$$\begin{aligned} x + y + z &= 0 \\ 2x + y + z &= 1 \\ 3x + 2y + z &= 1 \\ x + 2y + 3z &= 1 \\ 2x + 4y + 6z &= 2 \end{aligned} \quad (6)$$

From the above discussion, it is safe to say that the development of current methods for minimal over-constraint detection is far from being sufficient. The insufficiency has its origins in the inherent limitations of greedy methods. Again, two directions may be taken to solve the limitation or insufficiency. It can be seen from the examples that greedy methods gave the wrong detection results due to the wrong choice of the seed equation. For example, if the seed equations is set to equations E4, the greedy methods may give the right detection results. Therefore, one improvement direction is to have a preprocessing that finds that optimal seed equation. This problem is however a challenging issue for its own sake and remains an open issue. The other improvement direction, which is more desirable, is to put effort into the insights of the detection problem and to attain a solid mathematical modeling of this problem. The authors have presented a new, precise formulation of the minimal over-constraint detection problem in [30] to address Limitation II, following the latter direction.

### 3.4. Limitation III

The limitation this subsection discusses is the ineffectiveness of the current methods on maximal well-constrained part detection. Maximizing sizes of well-constrained parts is equivalent to minimizing under-constrained parts because under-constraint is described by DOFs between well-constrained parts. A well-constrained part of a GCS refers to a subset of the participating geometric entities whose induced constraint subsystem is well-constrained. An induced subsystem of a part is the subset of the constraints that are defined merely over the part. A well-constrained part $P$ is deemed as maximal if there is no another subset $P'$ of the participating geometric entities such that $P \subset P'$ and $P'$ is well-constrained.

Existing methods to solve the problem of maximal well-constrained part detection also use greedy methods [22,23], which leads to the limitation similar to Limitation II. The work flow of these methods is as follows. First, a geometric entity set $S$ is initialized with a seed geometric entity (randomly chosen); then, iterate through all the other geometric entities to check if the geometric entity form a well-constrained part with the geometric entities in *S*, and, if so, add the geometric entity to *S*; the resulting *S* gives a maximal well-constrained part. Repeat above procedures on the left geometric entities until no geometric entity is left. Similar to Limitation II, the initialization of the methods — the choice of seed geometric entities — has a significant effect on their effectiveness, and they may fail to detect the maximal well-constrained parts in a GCS.

An example of this limitation is given in Figs. 3 and 4. A crank model with two DOFs is shown in Fig. 3, and the two DOFs are a rotation of plane F2 about cylinder F4's axis and a rotation of plane F3 about cylinder F4's axis. Fig. 4a shows one detection result of using the greedy method, where the seed geometric entity for part I is plane F3, the seed geometric entity for part II is cylinder F7, and that for part III is plane F2. However, Fig. 4b shows another solution to the problem of maximal well-constrained part detection, with a larger size on the well-constrained part I.

This limitation, again, has its origins in the inherent limitations of greedy methods. In fact, the use of greedy algorithms does not give a formulation of the detection problem but represents an incomplete technical tool. To address the issue, we need to develop a mathematical modeling of the maximal requirement on sizes of well-constrained parts. The authors have presented a new, precise formulation of the maximal well-constraint detection problem in [30] to address Limitation III.

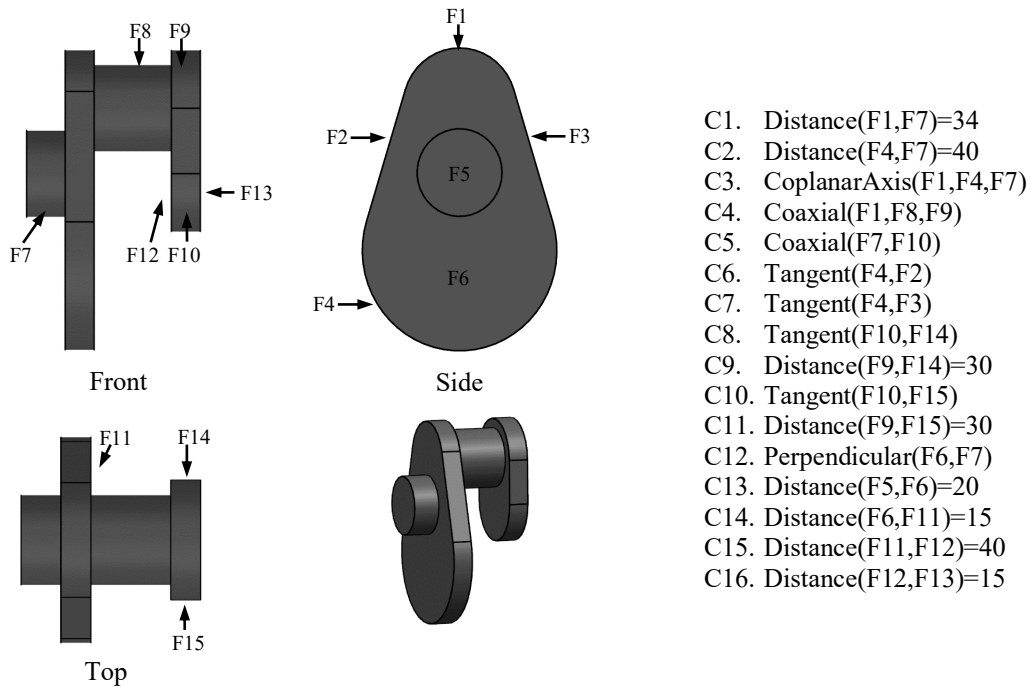

Figure 3  An under-constrained GCS example.

*3.5. Limitation IV*

This limitation is related to the third issue in geometric constraint solving — the optimal decomposition of a well-constrained GCS into the minimal well-constrained parts. It is not strictly necessary to have a VCM-based method to decompose a well-constrained GCS into the minimal well-constrained parts. The major advantage of VCM over the previous graph-based methods [4] is the ability to handle constraint dependency, but the setting of the third issue excludes such dependencies as the given GCS is well-constrained. There have been effective methods to accomplish the task of decomposing a well-constrained GCS into the minimal well-constrained parts, such as the graph-based methods [4].

For VCM to be self-contained and systematic, we want to have a VCM-based method to decompose a well-constrained GCS into the minimal well-constrained parts (although this may not have a practical value). To our knowledge, there is no existing work on this topic. This issue, as already noted, is secondary.

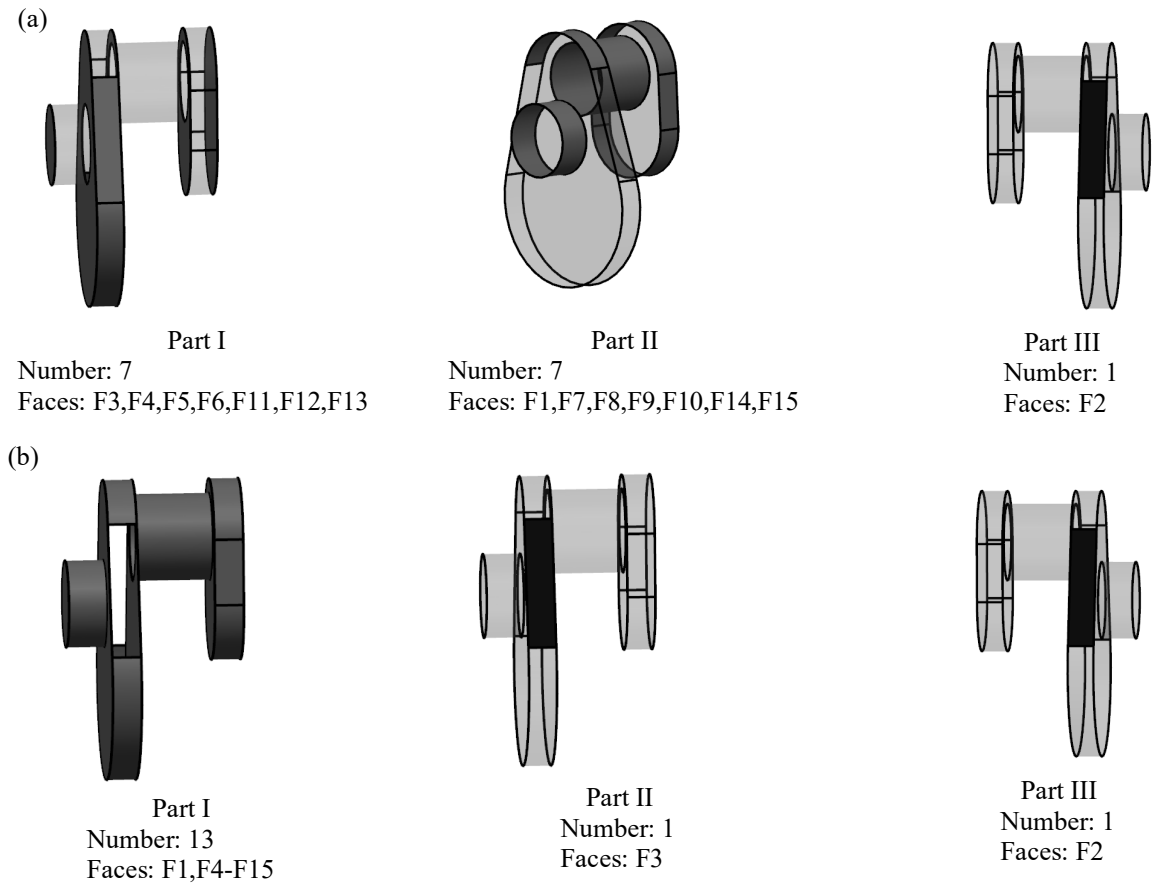

Figure 4   Results of maximal well-constrained part detection.

## 4. Conclusions

Limitations of VCM have been discussed in this work, under the context of geometric constraint solving. To have a systematic discussion of these limitations, we first outlined the three essential tasks/issues in geometric constraint solving, then introduced the basic idea of VCM in a brief fashion, and finally showed the capabilities and limitations of VCM using examples. The history of geometric constraint solving has also been briefly reviewed. With the limitations stated above, we expect that VCM can be perfected, starting from solving these limitations.

**Acknowledgments**

This work is a byproduct of the authors developing a new CAD modeling approach: variational direct modeling. The associated project was in part funded by the Natural Sciences and Engineering Research Council of Canada (NSERC). This financial support is greatly appreciated.

Technical Report, CAD/CAM/CAI Lab, MECH, UBC[2]  Coons S, and R. W. Mann. Computer-aided design related to the engineering design process. M.I.T. Electronic Systems Laboratory; 1960.

[3]  Shah JJ. Designing with parametric CAD: classification and comparison of construction techniques. Geometric Modelling, Springer; 2001, p. 53–68.

[4]  Hoffman CM, Lomonosov A, Sitharam M. Decomposition plans for geometric constraint systems, part I: performance measures for CAD. Journal of Symbolic Computation 2001;31:367–408.

[5]  Zou Q, Feng H-Y. Push-pull direct modeling of solid CAD models. Advances in Engineering Software 2019;127:59–69.

[6]  Zou Q, Feng H-Y. Push-pull direct CAD modeling with movable neighboring faces for preserving $G^1$ connections. ArXiv:190608455 2019.

[7]  Sutherland IE. Sketchpad a man-machine graphical communication system. Simulation 1964;2:R-3.

[8]  Bettig B, Hoffmann CM. Geometric constraint solving in parametric computer-aided design. Journal of Computing and Information Science in Engineering 2011;11:021001.

[9]  Dufourd J-F, Mathis P, Schreck P. Geometric construction by assembling solved subfigures. Artificial Intelligence 1998;99:73–119.

[10] Joan-Arinyo R, Soto A. A correct rule-based geometric constraint solver. Computers & Graphics 1997;21:599–609.

[11] Owen JC. Algebraic solution for geometry from dimensional constraints. Proceedings of the First ACM symposium on Solid Modeling Foundations and CAD/CAM Applications, 1991, p. 397–407.

[12] Bouma W, Fudos I, Hoffmann C, Cai J, Paige R. Geometric constraint solver. Computer-Aided Design 1995;27:487–501.

[13] Fudos I, Hoffmann CM. A graph-constructive approach to solving systems of geometric constraints. ACM Transactions on Graphics 1997;16:179–216.

[14] Gao X-S, Hoffmann CM, Yang W-Q. Solving spatial basic geometric constraint configurations with locus intersection. Proceedings of the Seventh ACM Symposium on Solid Modeling and Applications, 2002, p. 111–22.

[15] Bardord LA. A graphical, language-based editor for generic solid models represented by constraints. Cornell University, 1987.

[16] Serrano D. Constraint management in conceptual design. Massachusetts Institute of Technology, 1987.

[17] Ait-Aoudia S, Jegou R, Michelucci D. Reduction of constraint systems. Proceedings of Compugraphics, 1993, p. 83–92.

[18] Latham RS, Middleditch AE. Connectivity analysis: a tool for processing geometric constraints. Computer-Aided Design 1996;28:917–28.

[19] Hoffmann CM, Lomonosov A, Sitharam M. Finding solvable subsets of constraint graphs. Proceedings of International Conference on Principles and Practice of Constraint Programming, 1997, p. 463–77.

[20] Michelucci D, Foufou S. Geometric constraint solving: the witness configuration method. Computer Aided Design 2006;38:284–99.

[21] Graver JE, Servatius B, Servatius H. Combinatorial rigidity. American Mathematical Society; 1993.

[22] Thierry SEB, Schreck P, Michelucci D, Funfzig C, Génevaux JD. Extensions of the witness method to characterize under-, over- and well-constrained geometric constraint systems. Computer Aided Design 2011;43:1234–49.

[23] Foufou S, Michelucci D. Interrogating witnesses for geometric constraint solving. Information and Computation 2012;216:24–38.

[24] Kubicki A, Michelucci D, Foufou S. Witness computation for solving geometric constraint systems.
9